\documentclass[12pt]{iopart}
\usepackage{iopams} 
\usepackage{graphics}
\usepackage{pennames}
\usepackage{amssymb}
\usepackage{setstack}
\def\La{\Lambda}
\def\Al{\overline\Lambda}

\begin{document}
\title[Study of the $m_{\tt T}$\ spectra of strange particles at the SPS]
{Blast-wave analysis of strange particle $m_{\tt T}$\ spectra in Pb-Pb collisions at the SPS}
\author{G E Bruno for  
the NA57 Collaboration~\footnote[1]{For the full author list see 
Appendix ``Collaborations'' in this volume.}
}
\address{
Dipartimento~IA~di~Fisica~dell'Universit{\`a}~e~del~Politecnico~di~Bari~and~INFN,~Bari,~Italy\\
}
\begin{abstract}
The transverse mass spectra of high statistics,
high purity samples of \PKzS, \PgL, $\Xi$\ and  $\Omega$\ particles 
produced  in Pb-Pb collisions at SPS energy  
have been studied in the framework of the blast-wave model.    
The dependence of the freeze-out parameters on   
particle species and event centrality is discussed.  
Results at 40 $A$\ GeV/$c$\ are presented here for the first time.   
\end{abstract}
\vspace{-0.6cm}
%
%
%
\section{Introduction} 
The experimental programme with  
heavy-ion beams at CERN SPS is aimed at the study of 
the behaviour of 
hadronic matter 
under extreme conditions of temperature, pressure and energy density.  
Within this programme, NA57 is a dedicated experiment for the study of 
the production of strange and multi-strange particles 
in Pb-Pb and p-Be collisions at mid-rapidity~\cite{NA57proposal}.   
Results on strange particle yields, ratios and strangeness enhancements with 
respect to p-Be reactions 
are discussed in a separate contribution to this conference~\cite{EliaHQ04}.  

In this paper we present a study of the  
transverse mass ($m_{\tt T}=\sqrt{p_{\tt T}^2+m^2}$) 
spectra for  \PgL, \PgXm, \PgOm\ hyperons, 
their antiparticles and \PKzS\ 
measured in Pb-Pb collisions at 158 and 40 $A$\ GeV/$c$.  
The shapes of the $m_{\tt T}$\ spectra are expected to  
be determined both by the thermal motion of the 
particles and by a pressure-driven collective flow.  
To disentangle the two contributions  we rely on the {\it blast-wave} 
model~\cite{BlastRef,BlastRef2},  
which assumes cylindrical symmetry for an expanding fireball in local  
thermal equilibrium, testing different hypotheses on the transverse   
flow profile. 

\section{Data sample and analysis}
The results presented in this paper are based on the analysis of 
the full data sample collected in Pb-Pb collisions, consisting of 460 M events at 158 $A$\ GeV/$c$\ and
240 M  events at 40 $A$\ GeV/$c$.  
The selected sample of events corresponds to the most central 
53\% of the inelastic Pb-Pb cross-section.  
%
%
The data sample has been divided into five centrality classes (0,1,2,3 and 4,  
class 4 being the most central) according to the value of the charged particle 
multiplicity around central rapidity measured by a Silicon Microstrip Multiplicity 
Detector. 
The procedure for the measurement of the multiplicity distribution and 
the determination of the collision centrality for each class 
is described in reference~\cite{Multiplicity}. 
The fractions of the inelastic cross-section for the five classes, calculated 
assuming a total cross-section of 7.26 barn, are given  
in table~\ref{tab:centrality}.
\begin{table}[h]
\caption{Centrality ranges for the five classes.
\label{tab:centrality}}
\begin{center}
\begin{tabular}{llllll}
\hline
 Class &   $0$   &   $1$   &   $2$   &  $3$   &   $4$ \\ \hline
 $\sigma/\sigma_{inel}$\ \; (\%)   & 40 to 53 & 23 to 40 & 11 to 23& 4.5 to 11 & 0 to 4.5 \\
 \hline
\end{tabular}
\end{center}
\end{table}
\noindent

The experimental procedure for the determination of the $m_{\tt T}$\ distribution
is described in reference~\cite{BlastPaper}, where the results
at 158 $A$\ GeV/$c$\ are discussed in detail.

\section{Exponential fits of the transverse mass spectra at 158 $A$\ GeV/$c$}
\begin{figure}[t]
\centering
\resizebox{0.94\textwidth}{!}{%
\includegraphics{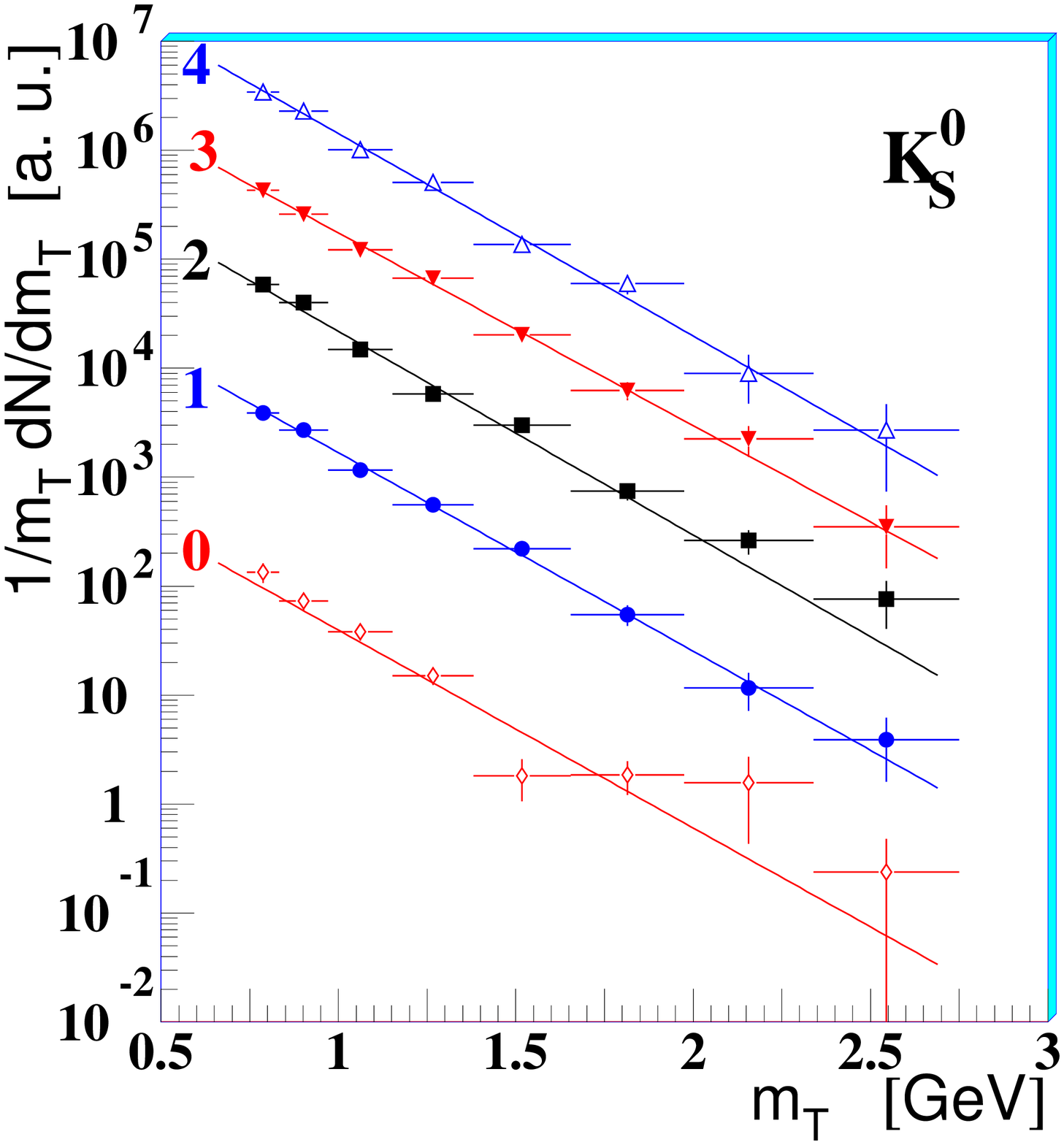}
\includegraphics{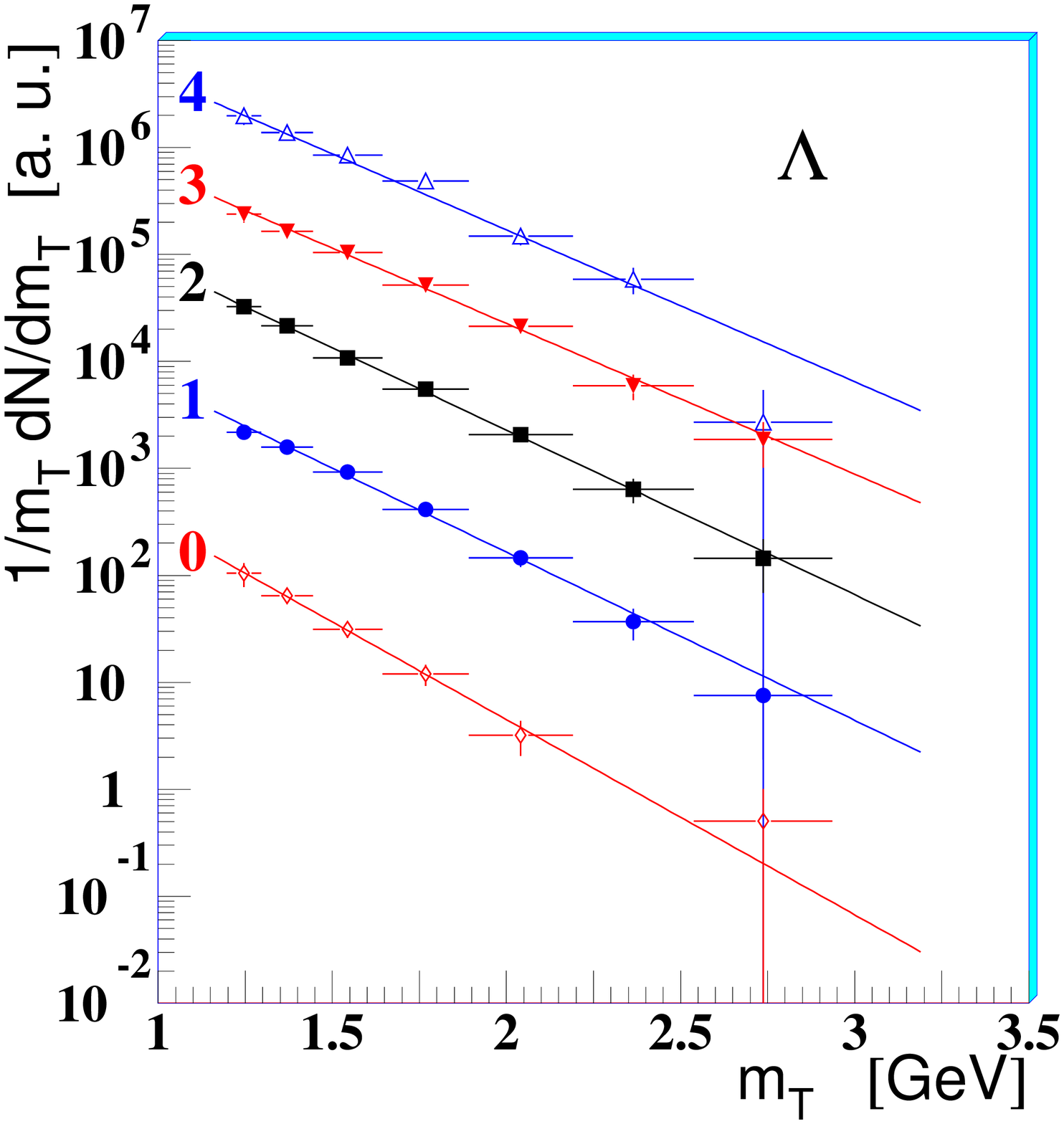}
\includegraphics{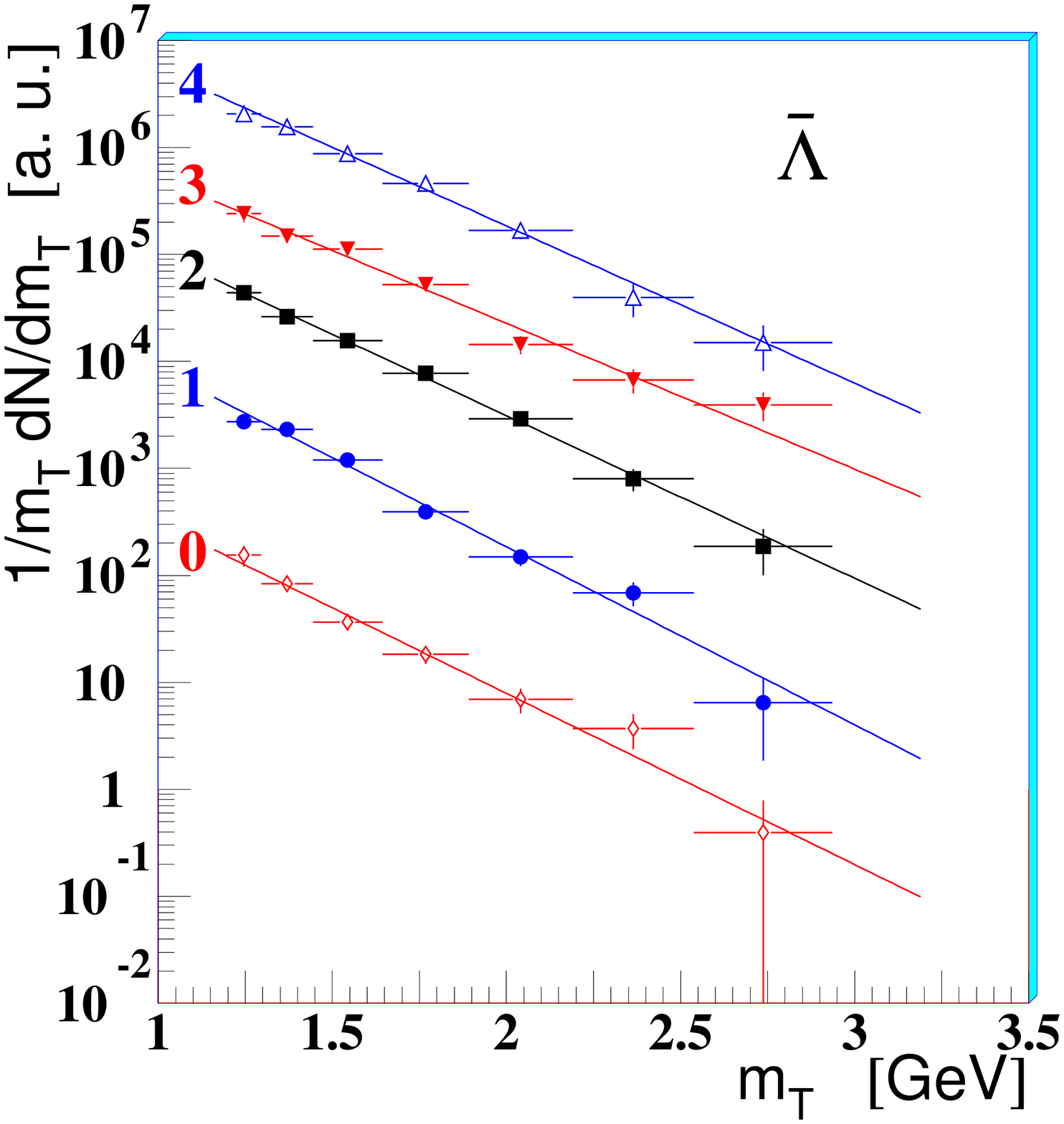}}\\
\resizebox{0.94\textwidth}{!}{%
\includegraphics{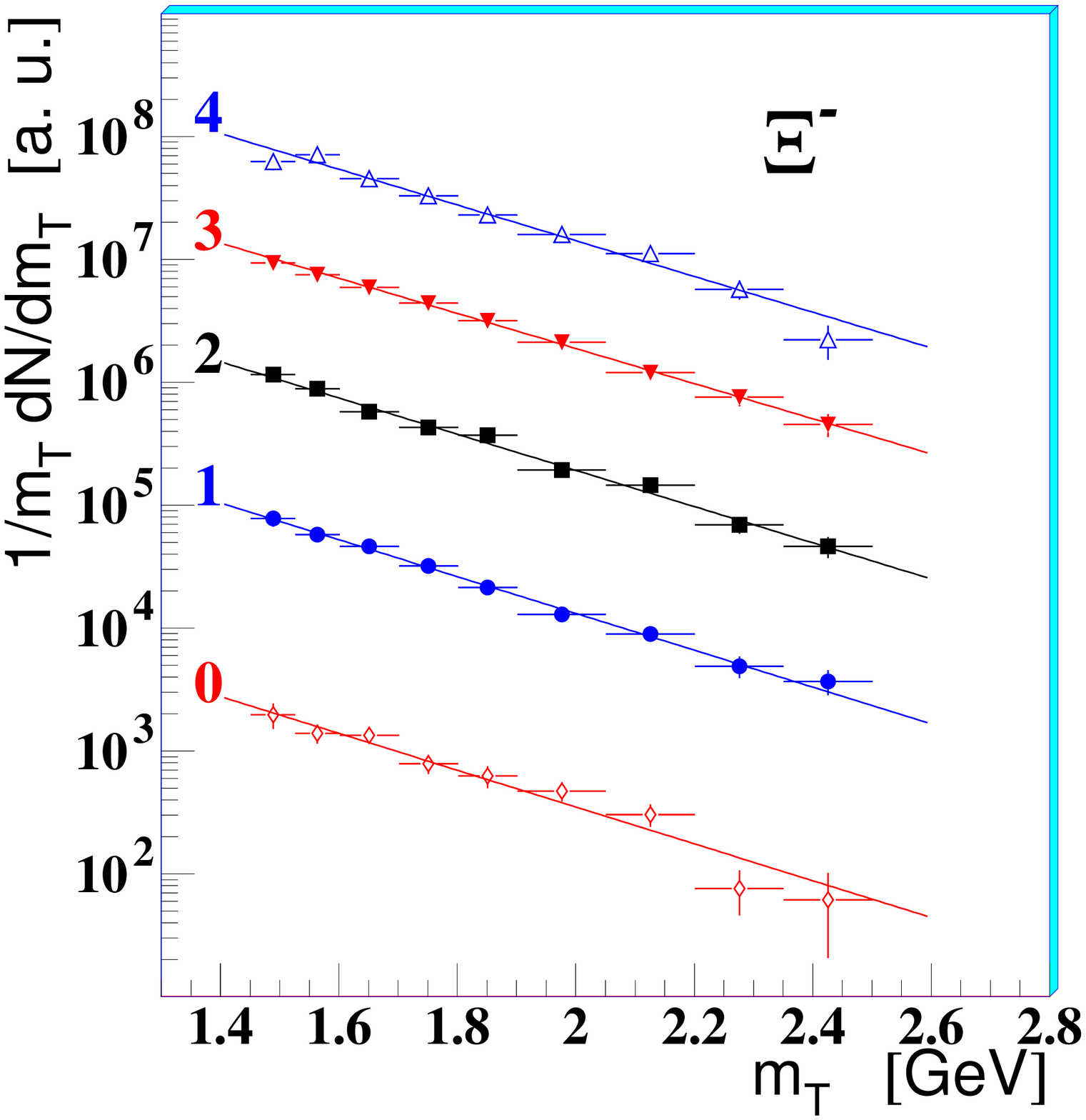}
\includegraphics{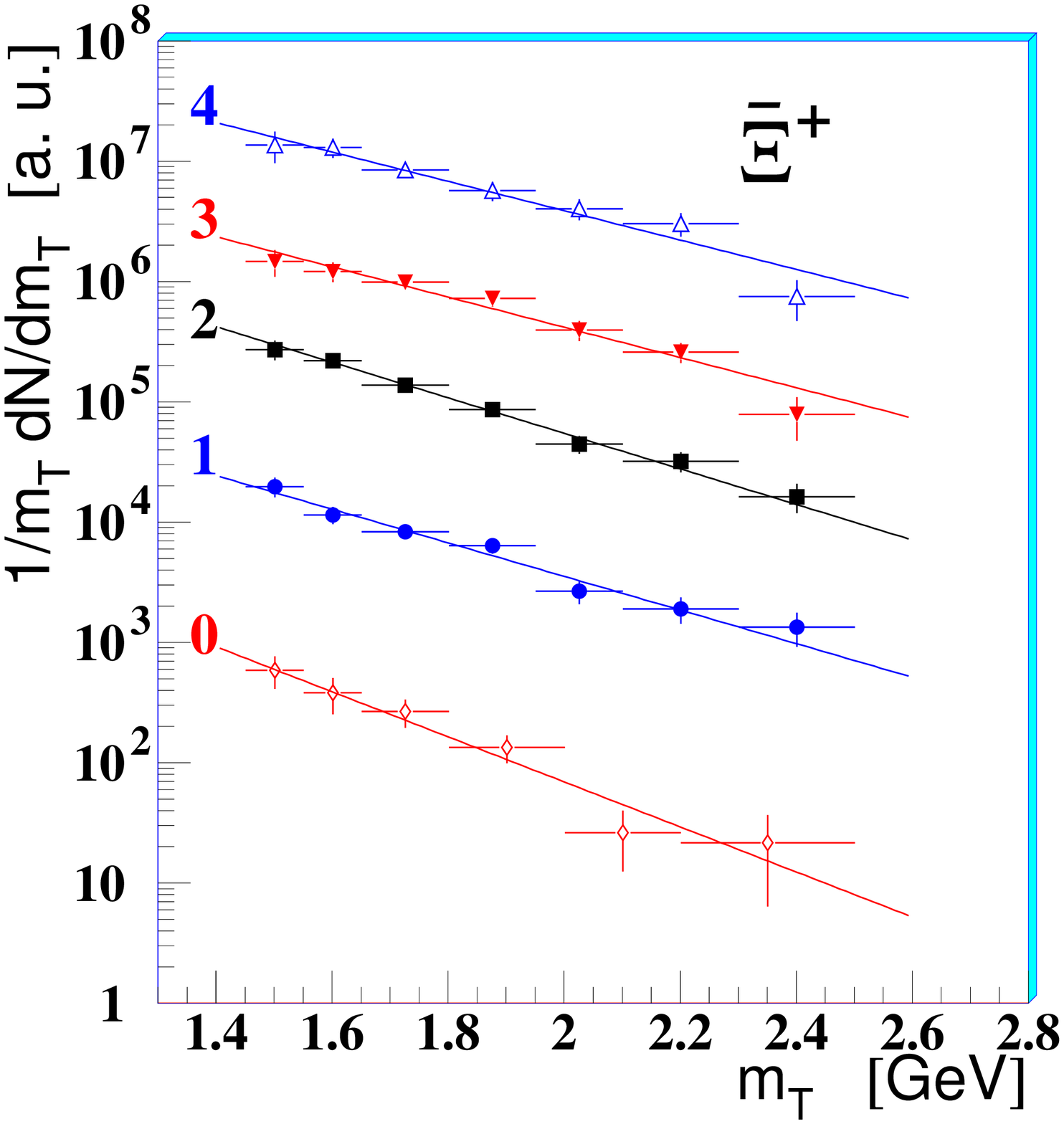}
\includegraphics{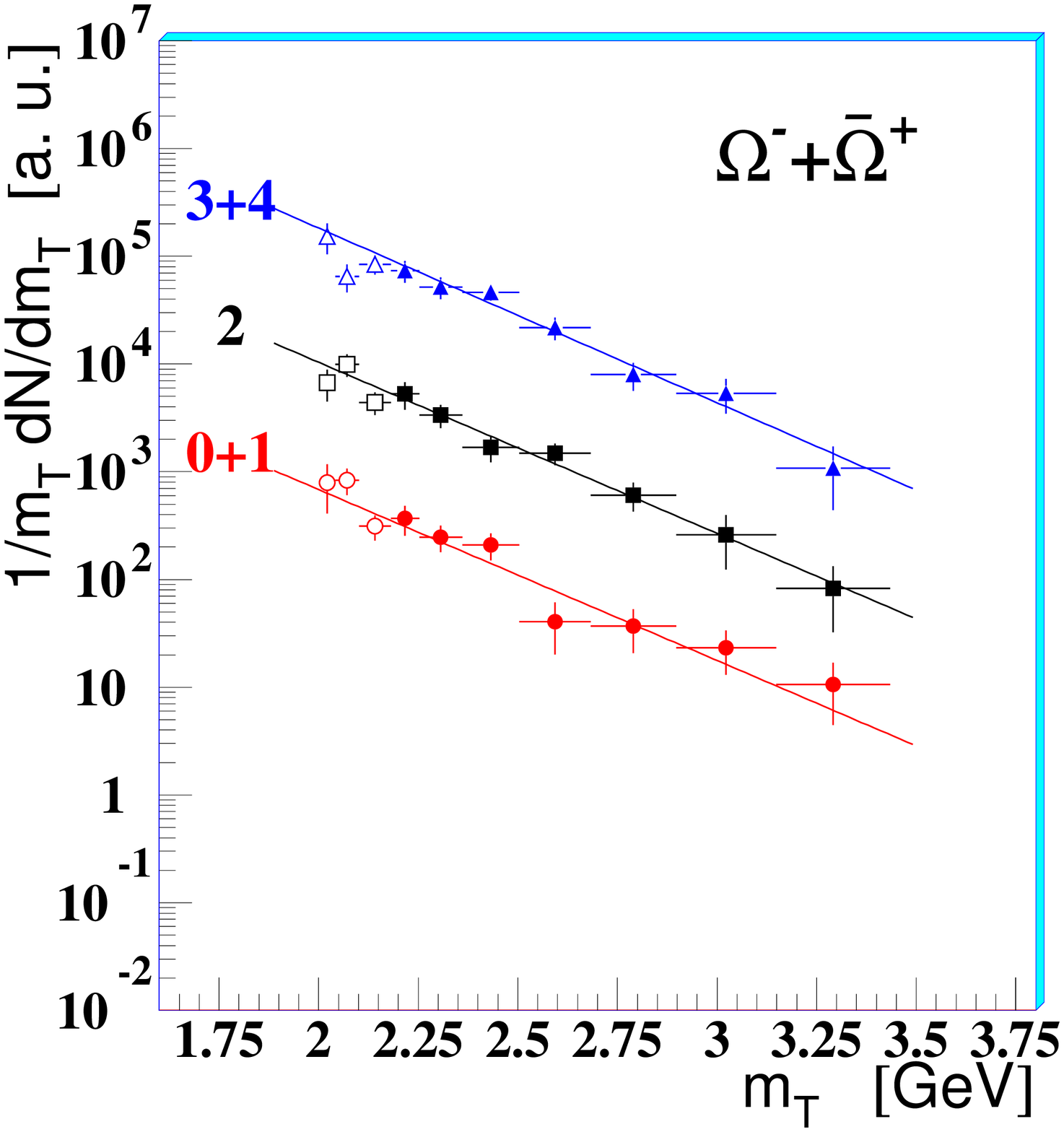}}
\caption{  Transverse mass spectra of the strange particles 
           in Pb-Pb collisions 
	   at 158 $A$\ GeV/$c$\ for
           the five centrality classes of table~\ref{tab:centrality}.
           For each species, class $4$\ is displayed uppermost and
           class $0$\ lowermost. 
	   }
\label{fig:msd_spectra}
\end{figure}

The inverse slope parameter $T_{app}$\ (``apparent temperature'')  
has been extracted by means of a maximum likelihood fit 
of the measured distribution of the invariant differential cross-section 
$\frac{d^2N}{m_{\tt T}\,dm_{\tt T} dy} $ 
to the formula 
$ 
f(y) \hspace{1mm} \exp\left(-\frac{m_{\tt T}}{T_{app}}\right) 
$\footnote{The rapidity distribution is assumed to be flat within our 
acceptance region corresponding to about half a unit of rapidity around mid-rapidity.}.  
The apparent temperature is interpreted as due to the  
thermal motion coupled with a collective transverse flow  
of the fireball components~\cite{BlastRef,BlastRef2}.    
The $1/m_{\tt T} \, dN/dm_{\tt T} $\ distributions are shown in figure~\ref{fig:msd_spectra} 
for the five centrality classes and the inverse slope parameters $T_{app}$\ are 
given in table~\ref{tab:InvMSD}.  
\begin{table}[h]
\caption{
         Inverse slopes (MeV) of the  
         $m_{\tt T}$\ distributions for the five Pb-Pb centrality 
	 classes ($0$,$4$), and 
	 for p-Be and p-Pb interactions~\cite{Slope-p} at 158 $A$\ GeV/$c$.  
	 Only statistical errors are shown. In Pb-Pb, systematic errors are 
	 estimated to be 10\% for all centralities.   
\label{tab:InvMSD}}
\begin{center}
\footnotesize{
\begin{tabular}{|c|c|c|cc|c|cc|} 
\hline
      &  p-Be   &   p-Pb    &    0      &    1      &    2      &    3    &    4   \\ \hline
\PKzS &$197\pm4$& $217\pm6$ & $239\pm15$ & $239\pm8$ & $233\pm7$ & $244\pm8$ & $234\pm9$ \\
\PgL  &$180\pm2$& $196\pm6$ & $237\pm19$ & $274\pm13$ & $282\pm12$ & $315\pm14$ & $305\pm15$ \\
\PagL &$157\pm2$& $183\pm11$ & $277\pm19$ & $264\pm11$ & $283\pm10$ & $313\pm14$ & $295\pm14$ \\
\PgXm &$202\pm13$&$235\pm14$ & $290\pm20$ & $290\pm11$ & $295\pm9 $ & $304\pm11$ & $299\pm12$ \\
\PagXp&$182\pm17$&$224\pm21$ & $232\pm29$ & $311\pm23$ & $294\pm18$ & $346\pm28$ & $356\pm31$ \\
\PgOm+
\PagOp&$169\pm40$& $334\pm99$ & \multicolumn{2}{c|}{$274\pm34$} & $274\pm28$ &
        \multicolumn{2}{|c|}{$268\pm23$} \\
\hline
\end{tabular}
}
\end{center}
\end{table}
An increase of 
$T_{app}$\ 
with  
centrality is observed in Pb-Pb  
for \PgL, \PagXp\  and possibly also for \PagL. 
Inverse slopes for p-Be and p-Pb collisions~\cite{Slope-p}  
are also given in table~\ref{tab:InvMSD}. In central  
and semi-central Pb-Pb collisions (i.e. classes 1 to 4) we observe 
baryon-antibaryon symmetry in the shapes of the spectra.  
Such a symmetry is not present in p-Be collisions.  
The similarity of baryon and antibaryon $m_{\tt T}$\ slopes observed in Pb-Pb  
suggests that strange baryons and antibaryons may be produced by similar   
mechanisms.   
\section{Blast-wave description of the spectra in Pb-Pb collisions}
%
The blast-wave model~\cite{BlastRef} predicts 
a double differential cross-section    
for particle $j$\   
of the form:  
\begin{equation}
\frac{d^2N_j}{m_{\tt T} dm_{\tt T} dy} 
    = \mathcal{A}_j  \int_0^{R_G}{ 
     m_{\tt T} K_1\left( \frac{m_{\tt T} \cosh \rho}{T} \right)
         I_0\left( \frac{p_{\tt T} \sinh \rho}{T} \right) r \, dr}
\label{eq:Blast}
\end{equation}
where $\rho(r)=\tanh^{-1} \beta_{\perp}(r)$\ is a transverse boost,   
$K_1$\ and $I_0$\ are  modified Bessel functions, $R_G$\ is the 
transverse geometric radius of the source at freeze-out 
and $\mathcal{A}_j$\ is a normalization constant.  
%
The transverse velocity field $\beta_{\perp}(r)$\ has been parametrized 
according to a power law: 
\begin{equation}
\beta_{\perp}(r) = \beta_S \left[ \frac{r}{R_G} \right]^{n}  
  \quad \quad \quad r \le R_G
\label{eq:profile}
\end{equation}  
With  
this type of profile the numerical value of $R_G$\ does not 
influence the shape of the spectra but just the absolute  normalization 
(i.e. the constant $\mathcal{A}_j$).  
%
The parameters which can be extracted from a fit of equation~\ref{eq:Blast} to 
the experimental spectra are thus the thermal freeze-out 
temperature $T$\ and the 
{\em surface} 
transverse flow velocity $\beta_S$. 
Assuming a uniform particle density, the 
latter can be connected to the {\em average} transverse flow 
velocity using the expression
$
<\beta_{\perp}> = \frac{2}{2+n}  \beta_S
$.   
\subsection{Global fit with different profiles}
The results 
of the fits  
with different profile hypotheses 
(i.e. different values of the exponent $n$) are given in table~\ref{tab:profile}.
\begin{table}[htb]
\caption{Results of the blast-wave model fit using different
         transverse flow profiles. The quoted errors are statistical.
         The systematic errors on the temperature and on the velocities
         are estimated to be about $10\%$\ and $3\%$, respectively,
         for all the four profiles.
\label{tab:profile}}
\begin{center}
\begin{tabular}{lllll}
\hline
                    & {\bf $n=0$}     & {\bf $n=1/2$}   & {\bf $n=1$}     & {\bf $n=2$}   \\ \hline
                          \multicolumn{5}{c}{\bf 158 $A$\ GeV/$c$ } \\
 $T$\ (MeV)         &  $158 \pm 6$    & $ 152\pm6 $     & $ 144\pm7 $     & $151 \pm 11$    \\
 $<\beta_{\perp}>$  & $0.396\pm0.015$ & $0.394\pm0.013$ & $0.381\pm0.013$ & $0.316\pm0.014$ \\
 $\chi^2/ndf     $  & $ 39.6/48 $     &  $ 36.9/48 $    & $ 37.2/48 $     & $68.0/48 $      \\
\hline
                          \multicolumn{5}{c}{\bf  40 $A$\ GeV/$c$ } \\
 $T$\ (MeV)         &  $128 \pm 4$    & $ 123\pm3 $     & $ 118\pm5 $     & $123 \pm 7$     \\
 $<\beta_{\perp}>$  & $0.432\pm0.012$ & $0.423\pm0.008$ & $0.398\pm0.010$ & $0.325\pm0.011$ \\
 $\chi^2/ndf     $  &   $ 62/34 $     &    $ 65/34 $    &   $ 79/34 $     &  $150/34 $      \\
\hline
\end{tabular}
\end{center}
\end{table}
\noindent
The values of $T$\ and $<\beta_\perp>$\ are found to be statistically
anti-correlated; the systematic errors on $T$\   
and $<\beta_{\perp}>$, instead, are correlated: 
they are estimated to be $10\%$\ and $3\%$, respectively, at both energies. 
The use of the three profiles $n$=0, 1/2 and 1 results in  
similar values of the freeze-out temperatures and of the  average transverse 
flow velocities. Profiles with $n\ge2$\  are disfavoured by the data.  

\begin{figure}[b]
\centering
\resizebox{1.00\textwidth}{!}{%
\includegraphics{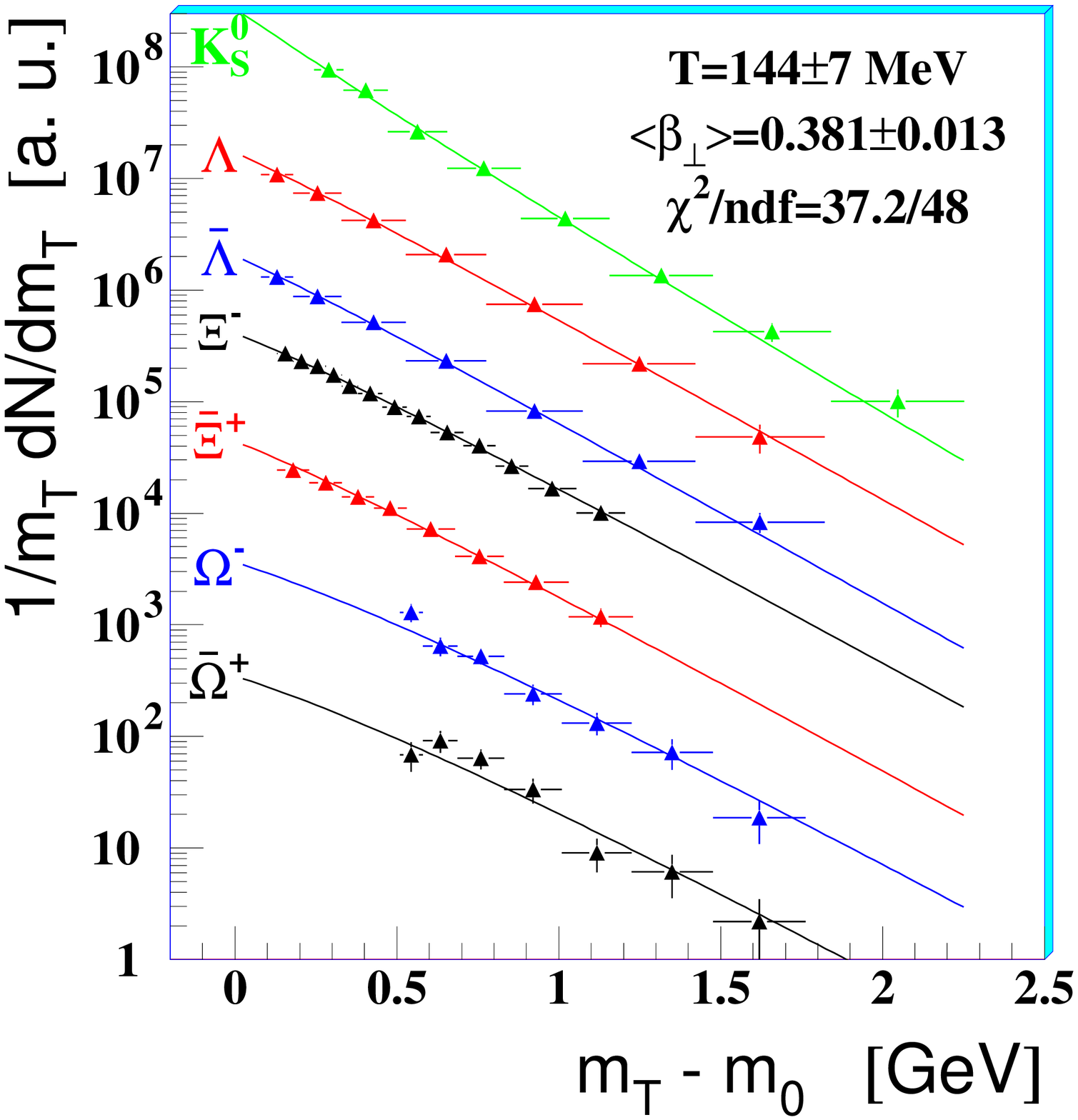}
\includegraphics{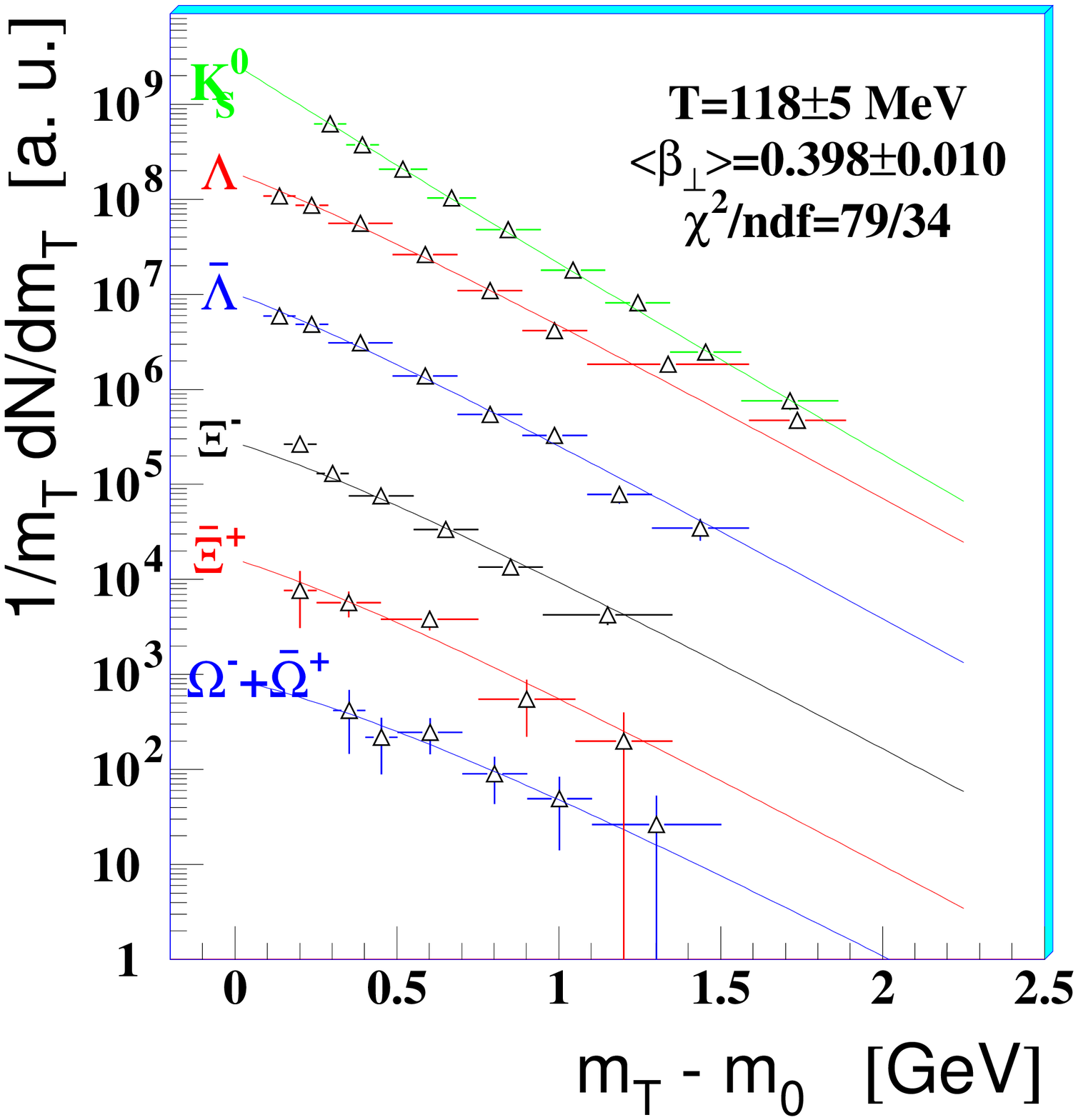}
\includegraphics{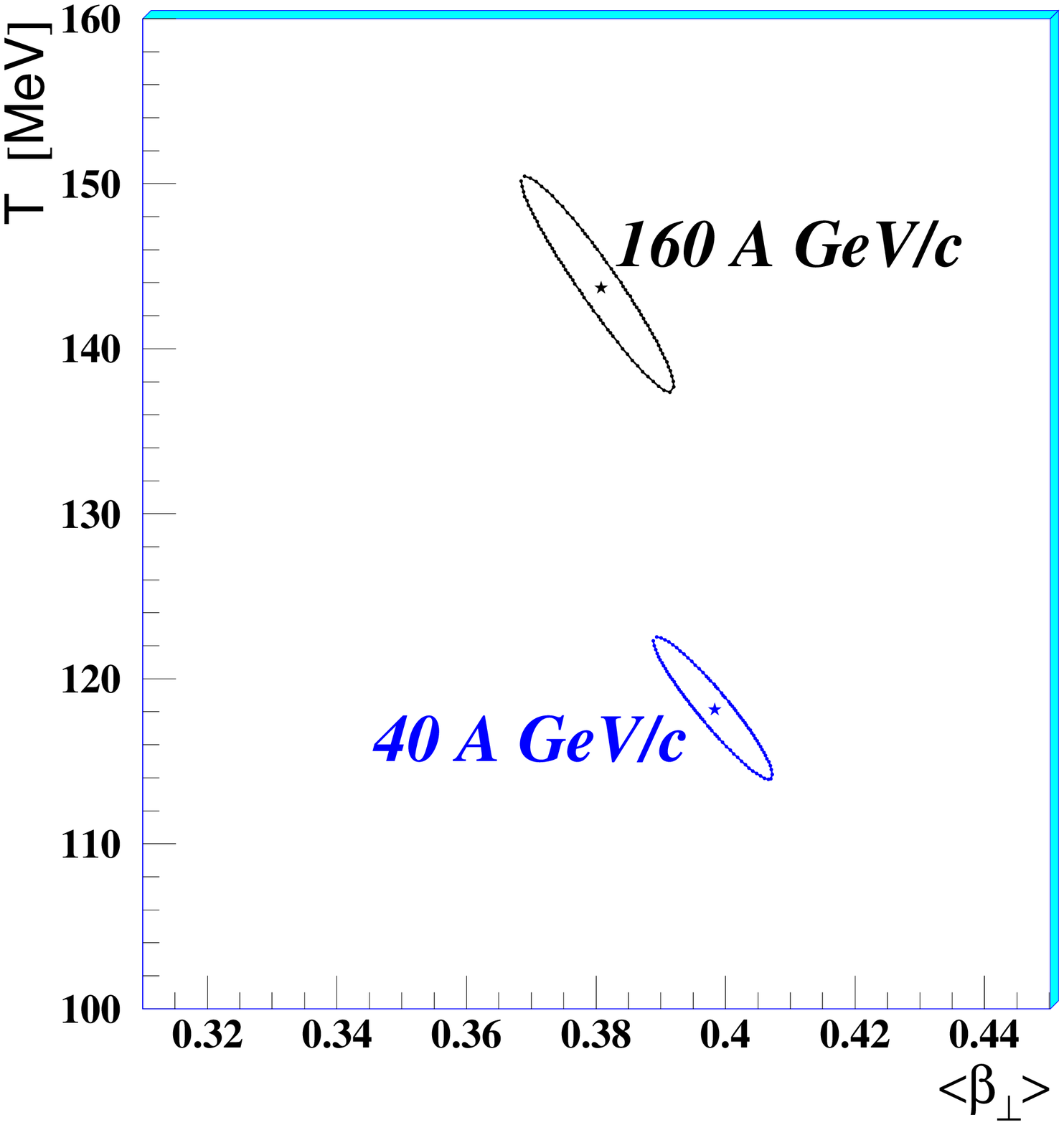}
}
\caption{ Blast-wave fits to the 
          transverse mass spectra of strange particles for the 53\%
          most central Pb-Pb cross-section at 158 (left) and at 40  
	  (middle) $A$\ GeV/$c$.  \\ 
	  Right: contour plots in the $<\beta_{\perp}>$--$T$\ 
	  plane at the 1$\sigma$\  confidence level.
          }
\label{fig:spettri_blast}
\end{figure}

In the following, a linear ($n=1$) $r$-dependence of the transverse 
flow velocity is used.  
The global fits of equation~\ref{eq:Blast} 
to the data points of all the measured strange particle spectra and the 
corresponding 1$\sigma$\ contour plots  
are shown in figure~\ref{fig:spettri_blast}.  
A lower thermal freeze-out temperature is measured at lower beam energy but  
the transverse flow velocities are found to be compatible in the errors.  
At 40 $A$\ GeV/$c$, a large contribution to the $\chi^2$\  comes from  
the $\Xi$ spectra: the possibility of an  
early freeze-out of multi-strange particles is discussed below. 
%

\subsection{Particles with/without quarks in common with the nucleon.}
The particles have been divided into two groups --- those which share 
valence quarks with the nucleons and those which do not  --- since 
it is known that the particles of the two groups may exhibit different 
production features. Results of separate blast-wave fits are given 
in table~\ref{tab:Blast1}. 
\begin{table}[h]
\caption{Thermal freeze-out temperature  
and average transverse flow velocity in the full centrality range.    
The first error is statistical, the second one systematic.
\label{tab:Blast1}}
\begin{center}
\begin{tabular}{lccc}
\hline
particles & ${\bf T}$ (MeV) &  $<{\bf \beta_\perp}>$ & $\chi^2/ndf$ \\ \hline\hline
   \multicolumn{4}{c}{\bf 158 $A$\ GeV/$c$ } \\
{\bf $K_S^0$}, {\bf $\La$},  {\bf $\Xi^-$} &
$146 \pm 8 \pm 14 $ & $ 0.376 \pm 0.015 \pm 0.012 $ & $ 18.1/23 $ \\ 
{\bf $\Al$}, {\bf $\overline\Xi^+$}, {\bf $\Omega^-$}, {\bf $\overline\Omega^+$} &
$ 130\pm28 \pm 14 $ & $ 0.403 \pm 0.032 \pm 0.012 $ & $ 18.5/23 $ \\ \hline
   \multicolumn{4}{c}{\bf 40 $A$\ GeV/$c$ } \\
{\bf $K_S^0$}, {\bf $\La$},  {\bf $\Xi^-$} &
$ 119 \pm 5 \pm 11 $ & $ 0.40 \pm 0.01\pm 0.01  $ & $ 56.6/18 $ \\ 
{\bf $\Al$}, {\bf $\overline\Xi^+$}, {\bf $\Omega^-$}, {\bf $\overline\Omega^+$} &
$ 80 \pm 19 \pm 11 $ & $ 0.45 \pm 0.03  \pm 0.01  $ & $ 14.8/18 $ \\
\hline
\end{tabular}
\end{center}
\end{table}
\noindent
The freeze-out conditions are compatible within two sigmas.   
Since the interaction cross-sections for
the particles of the two groups are quite different, this finding would suggest  
limited importance of final state interactions  
(i.e. a rapid thermal freeze-out) and  
 similar production mechanisms for the two groups.  
\subsection{Earlier freeze-out of multi-strange particles ?}
\begin{figure}[t]
\centering
\resizebox{0.90\textwidth}{!}{%
\includegraphics{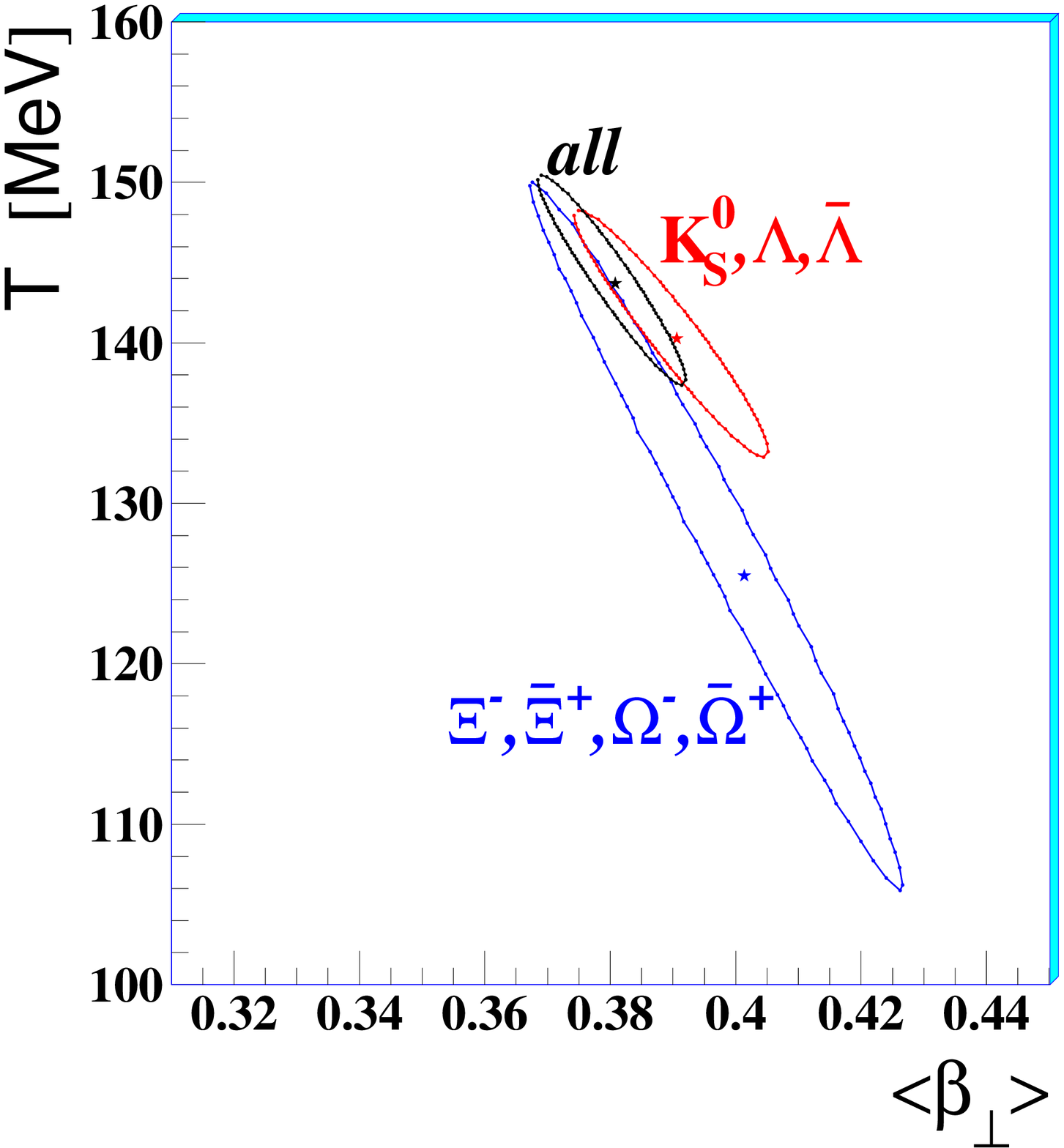}
\includegraphics{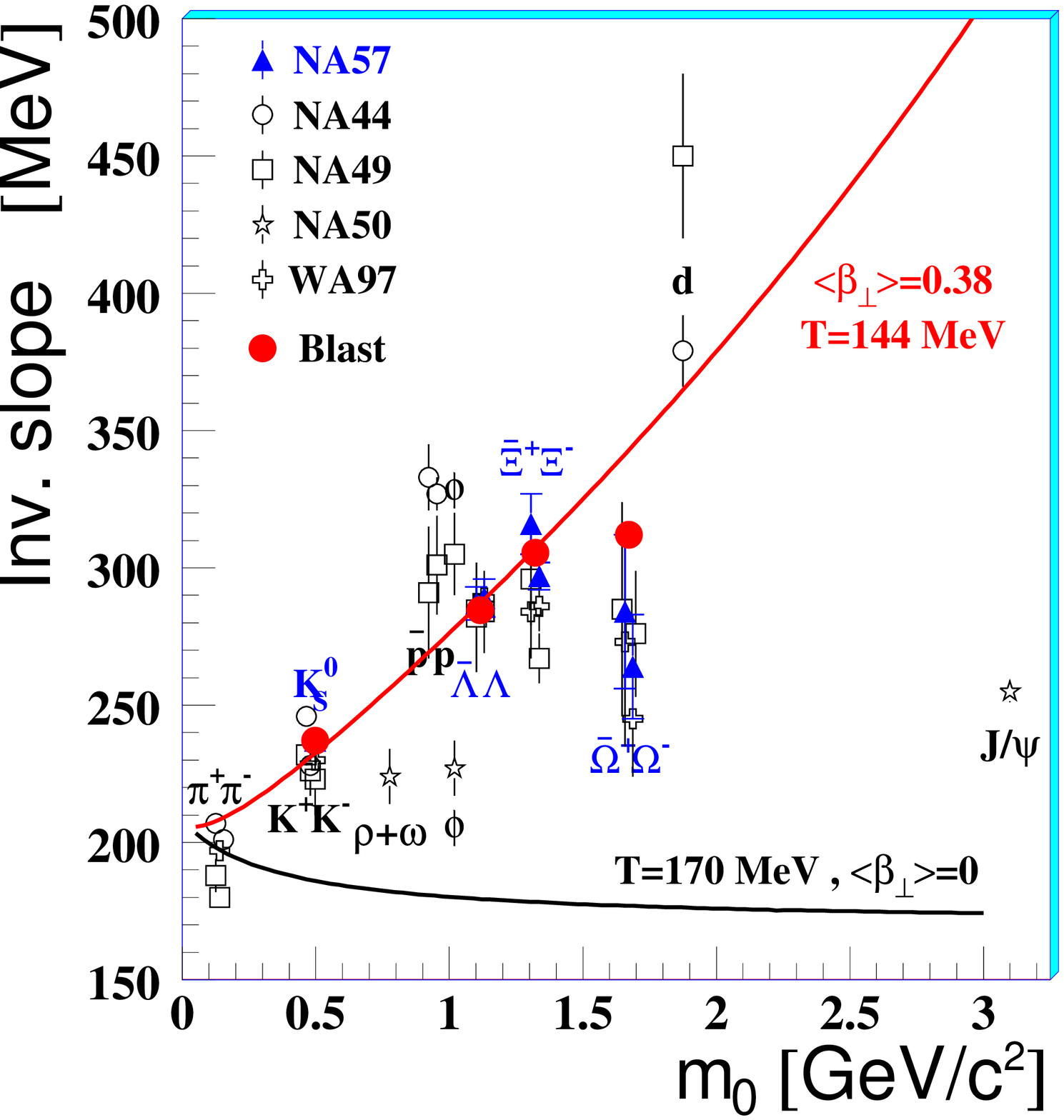}}\\
\resizebox{0.90\textwidth}{!}{%
\includegraphics{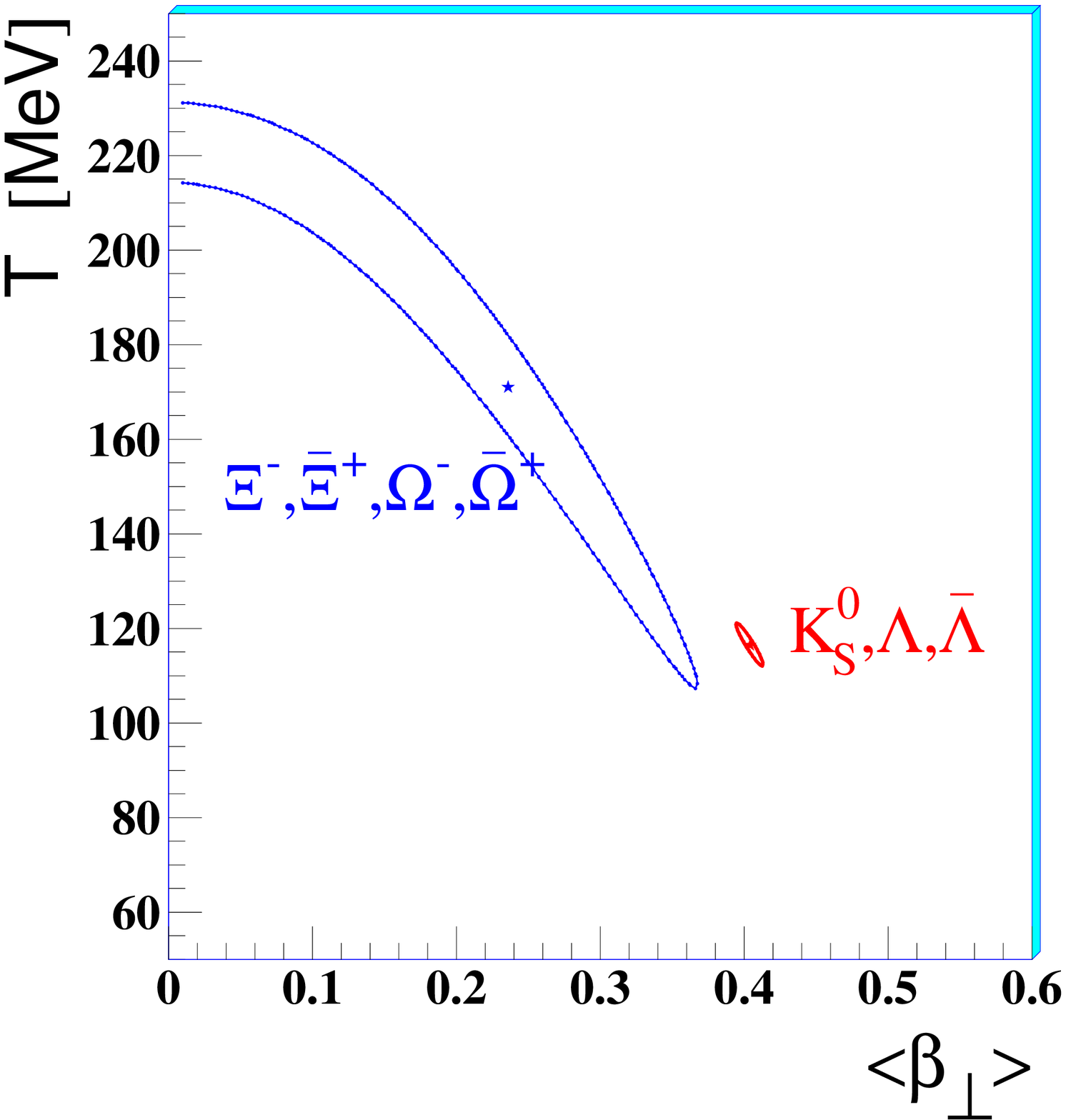}
\includegraphics{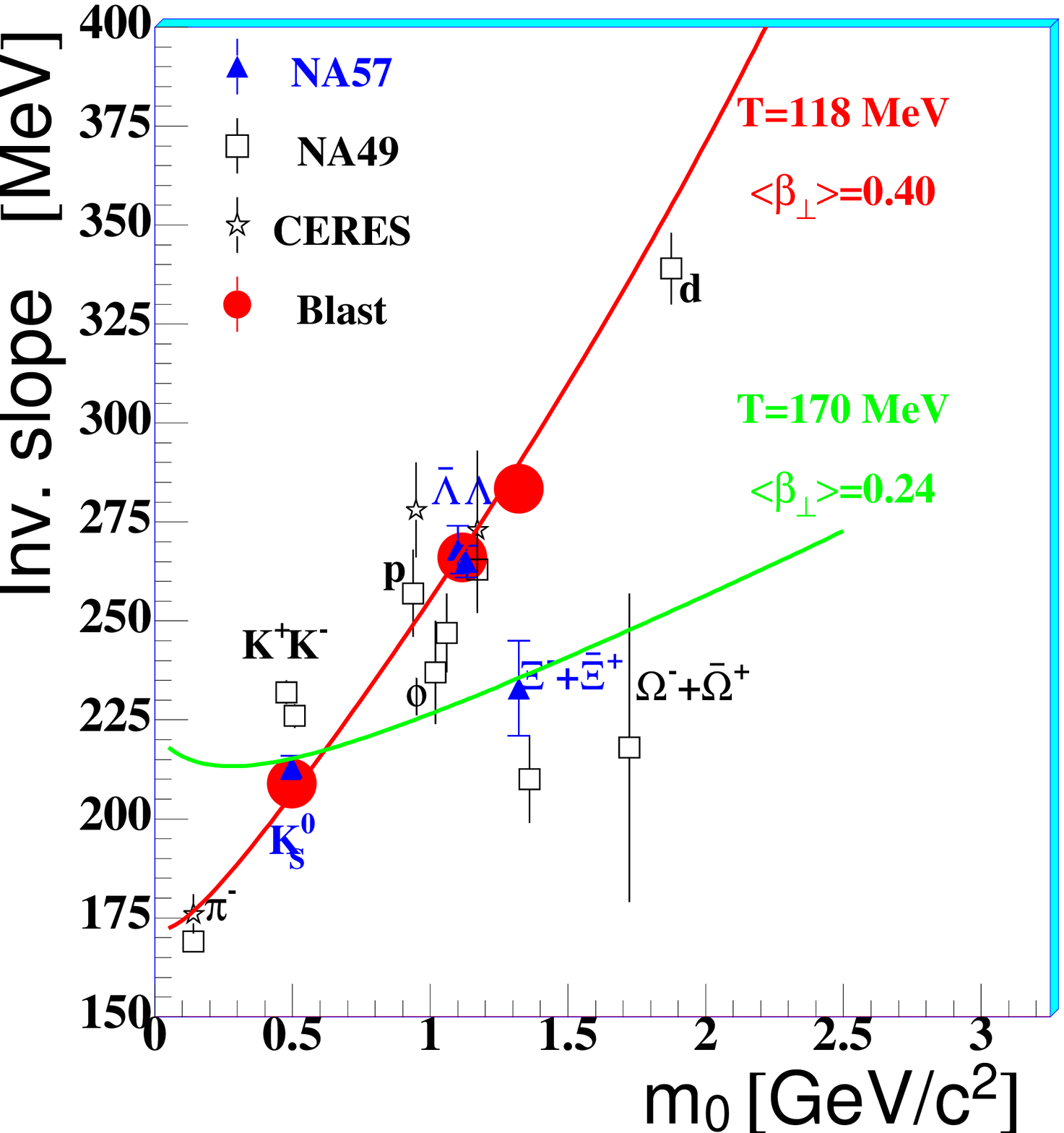}}
\caption{Top: 158 $A$\ GeV/$c$,  bottom: 40 $A$\ GeV/$c$.\\
         Left: the thermal freeze-out temperature versus the average transverse flow 
         velocity for blast-wave fits using a linear ($n=1$) velocity profile.   
         The 1$\sigma$\ contours are shown, with the markers indicating  
	 the optimal fit locations.  
	 Right: prediction of the blast-wave model for  
         inverse slopes (see text for details). 
	 }  
\label{fig:BlastPred}
\end{figure}

The 1$\sigma$\ contours of the separate blast-wave fits for singly  
and multiply strange particles are shown in figure~\ref{fig:BlastPred}. 

%
%
At 158 $A$\ GeV/$c$, the results of the fits for both
groups of particles are compatible with the result of the global fit
determination.
However, the fit for the multiply strange particles is statistically
dominated by the $\Xi$;  in fact the $\Xi+\Omega$\
contour remains essentially unchanged when fitting the $\Xi$\ alone.
For the $\Omega$, due to the lower statistics,  
it is not possible to extract significant values
for both freeze-out parameters
from its spectrum alone
(as can be done for the $\Xi$).
Any possible deviation for the $\Omega$\
from the freeze-out systematics extracted from the combined fit to the
$\PKzS$, $\Lambda$\ and $\Xi$\  spectra
can only be inferred from the integrated information of the
$\Omega$\ spectrum, i.e. from its inverse slope. 
In figure~\ref{fig:BlastPred} we plot a compilation of inverse slopes 
measured in Pb-Pb collisions,  superimposed to blast-wave model results.  
The full lines represent the inverse slope one would obtain by fitting  
an exponential to a ``blast--like'' $1/m_{\tt T} \, dN/dm_{\tt T}$\ 
distribution for a generic particle of mass $ m_{0} $,    
in the common range $ 0.05 < m_{\tt T} - m_{0}  < 1.50$\ GeV/$c^2$,   
for two different freeze-out conditions:   
absence of transverse flow ($<\beta_{\perp}>=0$) and our best fit determination.  
Since the inverse slope is a function of the $m_{\tt T} - m_{0}$\ 
range where the fit is performed, we have also computed the blast-wave 
inverse slopes of $\PKzS$, $\PgL$, $\Xi$\ and $\Omega$\  
spectra in the $m_{\tt T} - m_{0}$\ ranges of NA57  
(closed circles).  
The measured values of the inverse slope of the $\Omega$\ appear to deviate    
from the trend of the other strange particles.  

At 40 $A$\ GeV/$c$\ the same analysis suggests an early decoupling even  
of the $\Xi$\ with respect to the singly strange particles  
(see figure~\ref{fig:BlastPred}). 
\subsection{Centrality dependence}
\begin{figure}[t]
\centering
\resizebox{0.90\textwidth}{!}{%
\includegraphics{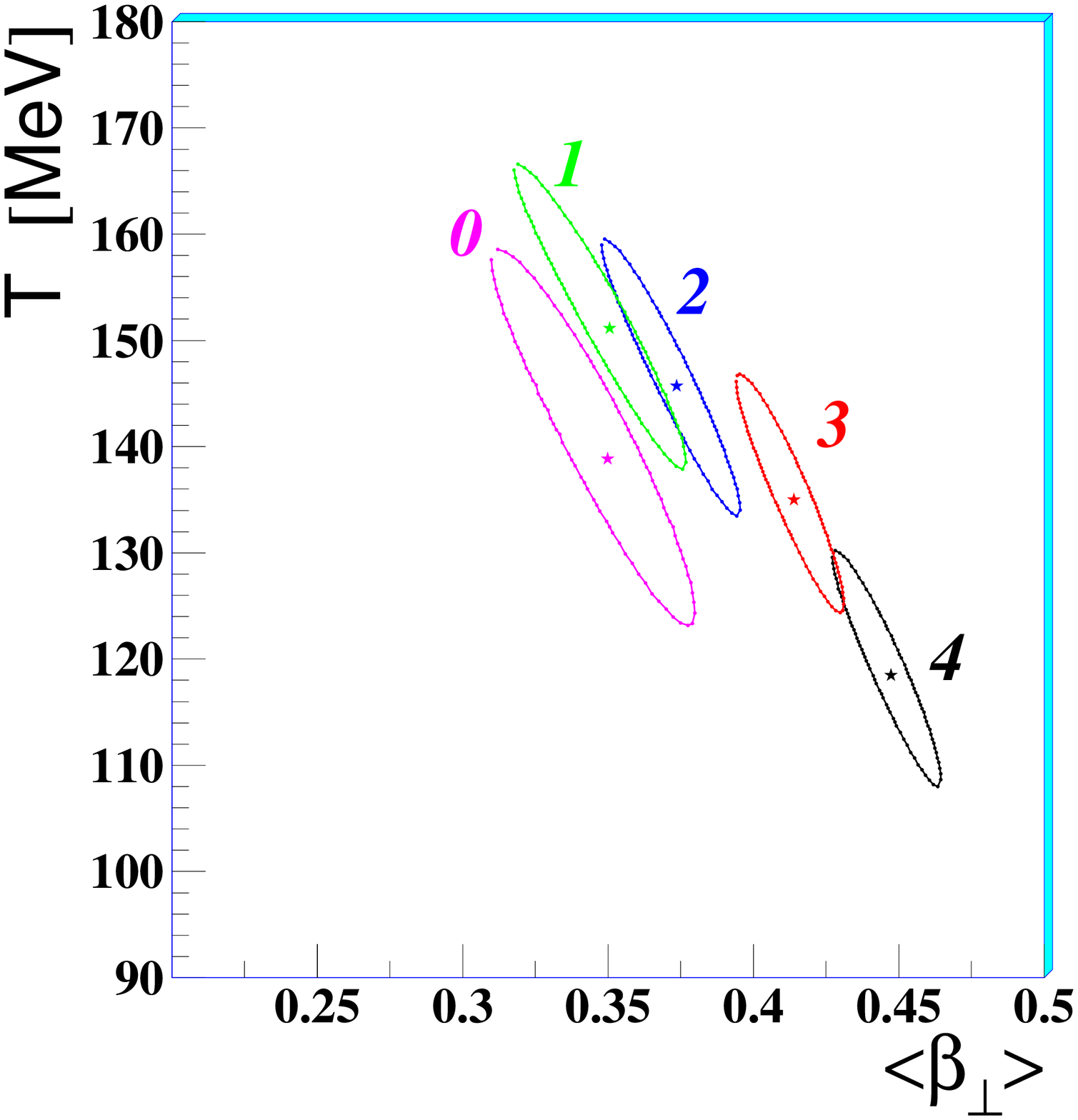}
\includegraphics{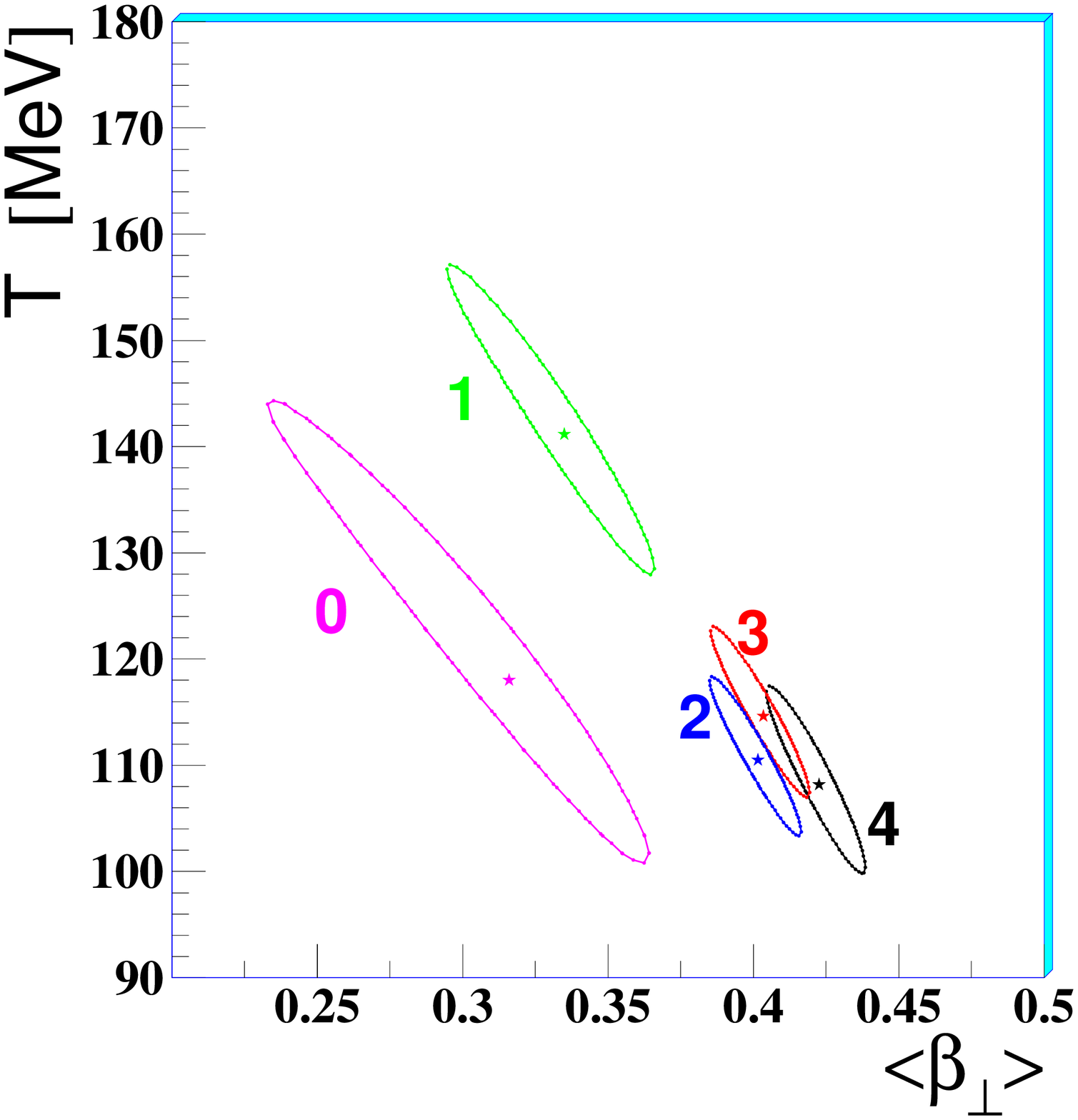}}
\caption{
         The 1$\sigma$ confidence level contours at 158 (left) and 40 (right) $A$ GeV/$c$.} 
\label{fig:cont_msd}
\end{figure}

In figure~\ref{fig:cont_msd} we show the $1\sigma$\ confidence level contours  
for each of the five centrality classes defined in table~\ref{tab:centrality}. 
The observed trend is as follows:  
the more central the collisions the larger the transverse collective 
flow and the lower the final thermal freeze-out temperature.  
Higher freeze-out temperatures for more peripheral collisions may be interpreted 
as the result of an earlier decoupling of the expanding system.  
Therefore, when trying to describe hydro-dynamically 
the experimental data measured for peripheral or semi-central collisions, 
one should employ higher values of the freeze-out temperature than the 120 MeV 
measured for central collisions.   
Indeed, in reference~\cite{NA45} the measured 
elliptic flow for Pb-Pb at 158 $A$\ GeV/$c$\ in the centrality range 
$\sigma/\sigma_{geo}=(13-26)\%$\ is close to that obtained with a hydro-dynamical 
evolution terminated at $T=160$\ MeV. 
In the range $\sigma/\sigma_{geo}=(11-23)\%$\  
we measure $T=146\pm17$\ MeV.

\section{Conclusions}
The analysis of the transverse mass spectra 
of strange particles in Pb-Pb collisions at SPS energies  
suggests  that after a central 
collision the system expands explosively and then it 
freezes-out when the temperature is of the order  
of 120 MeV, with an average transverse  flow   
velocity of about one half of the speed of light.  
Similar transverse flow velocities are measured at 40 and 158 $A$\ GeV/$c$\ 
but the freeze-out temperature is lower at the lower energy. 
The inverse slopes of multi-strange particles 
($\Omega$\ at 158 $A$\ GeV/$c$, $\Xi$\ and $\Omega$\ at 40 $A$\ GeV/$c$)   
appear to 
deviate from the values predicted by   
the blast-wave model tuned on singly-strange particles 
(\PKzS, \PgL\ and \PagL).    
Finally, the  results on the centrality dependence of the  
expansion dynamics  
indicate that with increasing centrality  
the transverse flow velocity increases and the freeze-out temperature 
decreases.  
%
%

\end{document}